\documentclass[aps,prl,twocolumn,superscriptaddress,groupedaddress]{revtex4}
\usepackage{graphicx}
\usepackage{color}
\usepackage{here}

\makeatletter
\g@addto@macro\bfseries{\boldmath}
\makeatother

\def \beq{\begin{equation}}
\def \eeq{\end{equation}}
\def\eqref#1{(\ref{#1})}
\def\bea{\begin{eqnarray}}
\def\eea{\end{eqnarray}}
\def\jpsi{J\kern-0.1em/\kern-0.1em\psi}

\def\URLtilde{\lower0.2em\hbox{$\tilde{\phantom{a}}$}}
\def\mycomm#1{\hfill\break\strut\kern-3em{\color{red}\tt ====> #1
\color{black}}\hfill\break}

\begin{document}
\setcounter{footnote}{1}

\title{First exotic hadron with open heavy flavor:  
$cs\bar u\bar d$ tetraquark}

\author{Marek Karliner} 
\email{marek@tauex.tau.ac.il}
\affiliation{School of Physics and Astronomy\\
University, Tel Aviv 69978, Israel}
\author{Jonathan L. Rosner}
\email{rosner@hep.uchicago.edu}
\affiliation{Enrico Fermi Institute and Department of Physics \\
University of Chicago, 5620 S. Ellis Avenue, Chicago, IL
60637, USA}

\date{August 21, 2020}

\begin{abstract}
The LHCb Collaboration has reported resonant activity in the channel $D^+ K^-$,
identifying two components: $X_0(2900)$ with $J^P = 0^+$ at $2866 {\pm} 7$ MeV,
$\Gamma_0=57{\pm} 13$ MeV and $X_1(2900)$ with $J^P = 1^-$ at $2904 {\pm} 7$ MeV,
$\Gamma_1=110{\pm} 12$ MeV.  We interpret the $X_0(2900)$ component as a $cs \bar
u\bar d$ isosinglet compact tetraquark, calculating its mass to be $2863 {\pm}
12$ MeV.  This is the first exotic hadron with open heavy flavor.  The 
analogous $bs\bar u\bar d$ tetraquark is predicted at $6213 {\pm} 12$ MeV.
We discuss possible interpretations of the heavier and wider $X_1(2900)$ state
and examine potential implications for other systems with two heavy quarks.
\end{abstract}
\smallskip

\maketitle

\sloppy
Very recently the LHCb Collaboration reported a narrow peak in the $D^+ K^-$
(+ c.c.) channel as seen in the decay $B^{\pm}\to D^+ D^- K^{\pm}$.  The peak has
been parametrized in terms of two Breit-Wigner resonances:
\bea \label{eqn0}
X_0(2900){:}
\kern 19em\strut
&&
\nonumber
\\
J^P{=}0^+,~M{=}2866{\pm}7~{\rm MeV},
~\Gamma_0{=}\phantom{1}57{\pm}13~{\rm MeV}~;
\kern 2em\strut
&&
\\ 
\strut\nonumber\\
X_1(2900){:}
\kern 19em\strut
&&
\nonumber
\\
J^P{{=}}1^-,~M{=}2904{\pm}5~{\rm MeV},
~\Gamma_1{=}110{\pm}12~{\rm MeV}~.
\kern 2em\strut
&&
\label{eqn1}
\eea
The statistical significance is $\gg 5\sigma$.
This is the first exotic hadron with open heavy flavor \cite{LHCbTalk:2020}. 

An obvious question is the interpretation of these states
and whether their structure and properties can be be elucidated
with our current understanding of nonperturbative QCD. 

We believe that regarding the lighter and narrower $0^+$ component
the answer is likely affirmative. We interpret this state
as a $cs\bar u\bar d$ compact $S$-wave tetraquark, with structure
completely analogous to the stable, deeply bound
$bb\bar u \bar d$ tetraquark expected well below $B B^*$ threshold.
The $I=0$ and $J^P=0^+$ isospin, spin and parity quantum numbers are 
robust consequences of this structure. That said, the $c$ and $s$ quarks being
significantly lighter than the $b$ quark results in a significant shift of the
tetraquark mass vs.\ the two-meson threshold.

We have approached this problem using two different methods, both based on
constituent quarks.  In the first, we note that different quark masses are
needed to describe mesons and baryons \cite{Lipkin:1978eh}, with baryonic
quarks approximately 55 MeV heavier than mesonic ones.  We used this method to
successfully anticipate \cite{Karliner:2014gca} the mass of the doubly charmed
baryon $\Xi_{cc}$ discovered by the LHCb experiment \cite{Aaij:2017ueg,note1},
and to predict the existence of a $bb \bar u \bar d$ tetraquark below threshold
for weak or electromagnetic decay \cite{Karliner:2017qjm}.  (See also 
Ref.~\cite{Eichten:2017ffp}.)

\begin{figure}[H]
\begin{center}
\includegraphics[width=0.47\textwidth]{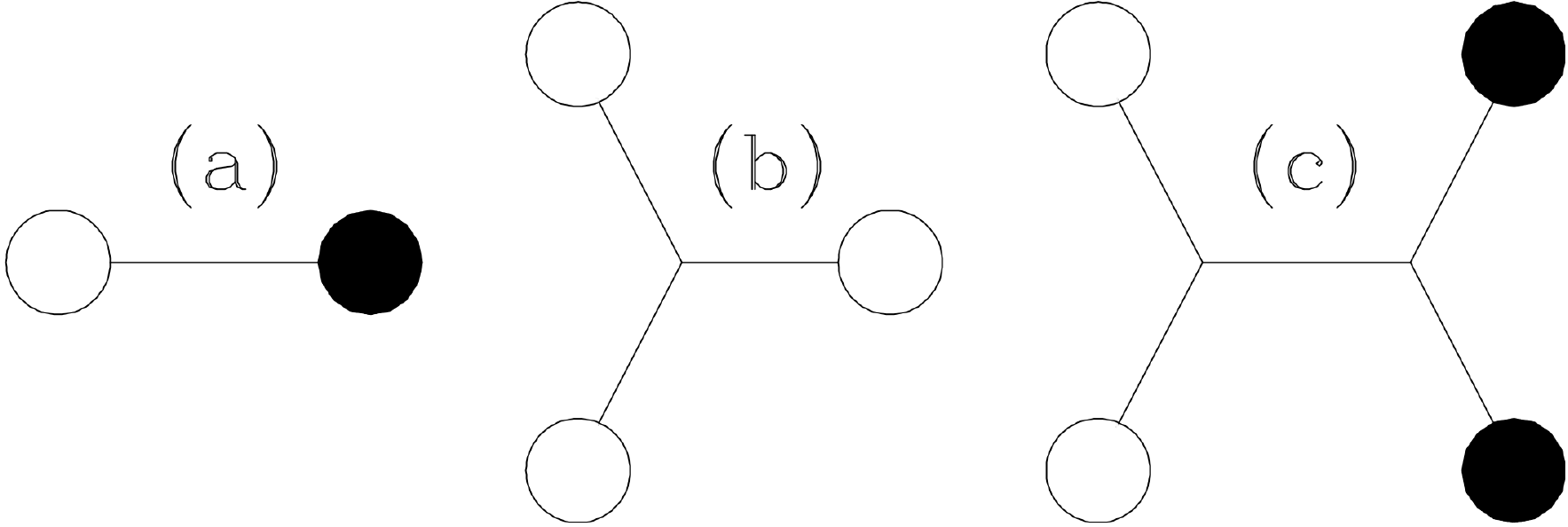}
\end{center}
\caption{\small%
QCD strings connecting quarks (open circles) and antiquarks
(filled circles).  (a) Quark-antiquark meson with one string and no
junctions; (b) Three-quark baryon with three strings and one junction;
(c) Baryonium (tetraquark) with five strings and two junctions.
\label{fig:jcts}}
\end{figure}

A second method, permitting the use of universal quark masses to describe
mesons and baryons, is to ascribe a mass contribution $S$ to each QCD string
junction \cite{Rossi:1977cy,Rossi:2016szw}, of which a meson has none, a baryon
has one, and a tetraquark has two, as illustrated in Fig.\ \ref{fig:jcts}.
%
%

\noindent
This method was used to predict masses of tetraquarks containing only heavy
quarks $c$ or $b$ \cite{Karliner:2016zzc}.

In the present Letter we compare the predictions of the string-junction picture
with the baryonic-quark picture for the mass of a ground state $J^P = 0^+$
tetraquark composed of $c s \bar u \bar d$, finding preference for the former
in light of the new LHCb result \cite{LHCbTalk:2020}.  The preferred
string-junction method is also applied to obtain the ground state mass of the
exotic tetraquark $bs \bar u \bar d$.  The two methods are then compared for
$bb \bar u \bar d$, $c c \bar u \bar d$, and $b c \bar u \bar d$ ground states,
and for $M(\Xi_{cc})$ originally calculated using baryonic quark masses
\cite{Karliner:2014gca}.

The LHCb Collaboration has reported two resonances in the $D^+ K^-$ (+c.c.)
channel around 2.9 GeV \cite{LHCbTalk:2020}: one with $J^P = 0^+$ and the
other broader one with $J^P = 1^-$ (see Eqs.\ (\ref{eqn0}),(\ref{eqn1}).)  
We can attempt to describe the $0^+$ state in either the baryonic-quark
picture or the string-junction (universal quark mass) picture, treating the
strange quark as heavy.

In a scheme with baryonic quarks, whose masses are labeled by superscripts $b$,
we follow the approach and the notation of Ref.~\cite{Karliner:2017qjm}
to obtain
\beq
M[T(cs \bar u \bar d,{\rm bar})] = 
m^b_c + m^b_s + m^b_{[ud]} + B(cs) + \Delta E_{HF}(cs)~,
\eeq
where $m^b_c = 1710.5$ MeV, $m^b_s = 538$ MeV, the mass of the $I=J=0$ diquark
is $m^b_{[ud]} = 576$ MeV, the binding energy of a $cs$ pair is $B(cs) = -35.0$
MeV, and the $cs$ hyperfine interaction accounts for $\Delta E_{HF}(cs) =
-35.4$ MeV (the last being taken from the footnote of Table IV of
Ref.~\cite{Karliner:2014gca}). The result is $M[T(cs \bar u \bar d,{\rm bar})]
 = 2754.1 \pm 12$ MeV, where the theoretical error is that encountered in Ref.\
\cite{Karliner:2014gca}.

In the string-junction picture, with universal-mass quarks, the relevant
parameters are $S = 165.1$ MeV, $m_s = 482.2$ MeV, $m_c = 1655.6$ MeV.
One can combine $m^{(b)}_c + m^{(b)}_{[ud]}=M(\Lambda_c)$ in either calculation,
so the only difference between the two calculations is one string-junction
term [a second is contained in $M(\Lambda_c)$] and the difference in strange
quark masses:
\bea
M[T(cs \bar u \bar d,{\rm str})] - M[T(cs \bar u \bar d,{\rm bar})] =
S + m_s - m^b_s &=& 
\nonumber
\\
= (165.1 + 482.2 - 538)~{\rm MeV} = 109.3~{\rm MeV}~,
\eea
or
\beq \label{eqn0th}
M[T(cs \bar u \bar d,{\rm str})] = (2863.4 \pm 12)~{\rm MeV}~.
\eeq
where the theoretical error is that encountered in fits to light-quark hadrons
\cite{Karliner:2016zzc}.
This is within 3 MeV of the experimental central value of the $0^+$ mass
(\ref{eqn0}).  A clear preference for the string-junction (universal quark mass)
picture emerges.

The state $0^+$ state in Eq.\ (\ref{eqn0th}) has a hyperfine $J^P = 1^+$
partner in which the $cs$ $0^+$ pair is replaced by a $1^+$, with 
$\Delta E_{HF}(cs) = {+}17.7$ MeV.  The mass of the state is 2916.5 MeV.  As
it is of unnatural parity, it cannot decay into $DK$, so cannot account for
the state (\ref{eqn1}).  The activity in the Dalitz plot giving rise to that
state must be due to another source.  One possibility is a consequence of
rescattering $DK \to D^* K^*$ (e.g., via pion exchange), whose threshold is at
2.9 GeV \cite{LHCbTalk:2020}.  

The string-junction picture may also be applied to the exotic configuration
$bs \bar u \bar d$.  The terms contributing to its mass are $m_b$ plus
an $I = J =0$ $\bar u \bar d$ ``good quark'' mass $m_{[ud]}$
lumped into $M(\Lambda_b) = 5619.5$ MeV;
a strange quark mass $m_s = 482.2$ MeV; a string junction term $S= 165.1$ MeV;
and a binding energy $B(bs) = -41.8$ MeV and $bs$ hyperfine term of $-12$ MeV,
both taken from \cite{Karliner:2014gca}.  The result is the prediction
\beq
M[T(bs \bar u \bar d,0^+)] = (6213 \pm 12)~{\rm MeV}.
\label{bsud:mass}
\eeq
It is interesting that this is very close to the $B^*K^*$ threshold
at 6216 MeV, while the $0^+$ $(cs\bar u\bar d)$ state in the charm sector 
at 2866 MeV is close to the $D^*K^*$ threshold at 2902 MeV.
\noindent
$T(bs\bar u\bar d)$ is approximately 440 MeV above the $BK$ threshold. 
It should be seen in 
\bea
T(bs \bar u \bar d) &\to& \bar B^0 K^- 
\\
{\rm and}\phantom{aaaaaaaaa}&& \nonumber
\\
T(bs \bar u \bar d) &\to& B^- K^0~.
\eea
The first mode is preferable, because it avoids the $s$\, vs. $\bar
s$ ambiguity associated with neutral kaons.
In principle it should be possible to observe this state in LHCb and
perhaps in other LHC experiments.

The contributions to the mass of the lightest tetraquark $T(bb\bar u\bar d)$
with two bottom quarks and $J^P=1^+$ are listed in Table \ref{tab:bbud}.  The
notation and the numerical values of the parameters are the same as in Table
VI and Table IX of Ref.\ \cite{Karliner:2014gca}.  Theoretical errors reflect
deviations from experiment of fits to light-quark states.

\begin{table}
\caption{\small%
Contributions to the mass of the lightest tetraquark 
$T(bb\bar u\bar d)$ with two bottom quarks and $J^P=1^+$.  Baryonic quarks as
used in Ref.\ \cite{Karliner:2017qjm} (with superscript $b$); universal quarks
as used in Ref.\ \cite{Karliner:2016zzc} (no superscript).  $q$ denotes $u$
or $d$ quark; isospin breaking ignored.
\label{tab:bbud}}
\begin{center}
\begin{tabular}{c r c r} \hline \hline
\multicolumn{2}{c}{Baryonic quarks} & \multicolumn{2}{c}{Universal quarks} \\
Contribution & Value (MeV)  & Contribution & Value (MeV) \\ \hline
    --   &   --    &  $2S$  &  330.2 \\ 
$2m^b_b$ & 10087.0 & $2m_b$ & 9977.2 \\
$2m^b_q$ &  726.0  & $2m_q$ &  617.0 \\
$a_{bb}/(m^b_b)^2$ & 7.8 & $a_{bb}/(m_b)^2$ & 7.8 \\
${-}3a/(m^b_q)^2$ & ${-}150.0$ & ${-}3a/(m_q)^2$ & ${-}151.2$ \\
$bb$ binding & ${-}281.4$ & $bb$ binding & ${-}266.1$ \\
Total &$10389.4\pm12$&Total &$10514.9\pm12$  \\ \hline \hline
\end{tabular}
\end{center}
\end{table}

The mass of the ground $bb \bar u \bar d$ $J^P = 1^+$ state is 125.5 higher
for universal quarks with two string junctions, 
but is still 89 MeV below the $B^-\bar B^{*0}$ threshold
and 44 MeV below threshold for decay to $B^-\bar B^0 \gamma$.

The calculation of the masses of the lightest \,$cc\bar u\bar d$\,
tetraquark masses proceeds analogously to $bb\bar u\bar d$.
In Table~\ref{tab:ccud} we provide the corresponding
contributions to the $cc\bar u\bar d$ mass.

The mass of $T(c c \bar u \bar d)$ 
with universal quarks and two string junctions is 117.8 MeV
higher than that with baryonic quarks, well above threshold for decay to a $D
D^*$ pair.

In Table~\ref{tab:bcud} we provide the corresponding contributions to the
$bc\bar u\bar d$ tetraquark mass.  The mass of $T(bc \bar u \bar d)$ 
with universal quarks and two string junctions is 121.6 MeV
higher than that with baryonic quarks.  Whereas in the baryonic-quark picture
this value was just below $\bar B D$ threshold, the value in the
universal-quark picture is well above threshold.

\begin{table}
\caption{\small Contributions to the mass of the lightest tetraquark 
$T(cc\bar u\bar d)$ with two charmed quarks and $J^P=1^+$.
\label{tab:ccud}}
\begin{center}
\begin{tabular}{c r c r} \hline \hline
\multicolumn{2}{c}{Baryonic quarks} & \multicolumn{2}{c}{Universal quarks} \\
Contribution & Value (MeV) & Contribution & Value (MeV) \\ \hline
   --    &   --   &  $2S$  &  330.2 \\
$2m^b_c$ & 3421.0 & $2m_c$ & 3311.2 \\
$2m^b_q$ & 726.0  & $2m_q$ &  617.0 \\
$a_{cc}/(m^b_c)^2$ & 14.2  & $a_{cc}/(m_c)^2$ & 14.2  \\
${-}3a/(m^b_q)^2$ & ${-}150.0$ & ${-}3a/(m_q)^2$ & ${-}151.2$ \\
$cc$ binding & ${-}129.0$  & $cc$ binding & {-}121.3 \\
Total & $3882.2 \pm 12$ & Total & $4000.1 \pm 12$\\ \hline \hline
\end{tabular}
\end{center}
\end{table}

We compare in Table \ref{tab:compxi} the predictions for $M(\Xi_{cc})$
in the baryonic and universal quark mass schemes.
The predictions of the two schemes are almost the same, and both compatible
with the observed value of $(3621.55 \pm 0.23 \pm 0.30)$ MeV
\cite{Aaij:2019uaz}.  The shift of
three quark masses each by about 55 MeV is compensated by the $S$ term, and
the only remaining difference is about 8 MeV in the $cc$ binding term.
Hence it is really only when one gets to configurations with two or more
string junctions that a distinction emerges.

\begin{table}
\caption{\small Contributions to the mass of the lightest tetraquark 
$T(bc\bar u\bar d)$ with one bottom and one charmed quark and $J^P=0^+$.
\label{tab:bcud}}
\begin{center}
\begin{tabular}{c r c r} \hline \hline
\multicolumn{2}{c}{Baryonic quarks} & \multicolumn{2}{c}{Universal quarks} \\
Contribution & Value (MeV) & Contribution & Value (MeV) \\ \hline
     --       &   --   &    $2S$   &  330.2 \\
$m_b^b+m^b_c$ & 6754.0 & $m_b+m_c$ & 6644.2 \\
$  2m^b_q$    &  726.0 &   $2m_q$  &  617.0 \\
${-}3a_{bc}/(m_b^b m^b_c)$ & ${-}$25.5 & ${-}3a_{bc}/(m_b m^c)$ & ${-}$25.5 \\
${-}3a/(m^b_q)^2$ & ${-}150.0$ & ${-}3a/(m_q)^2$ & ${-}151.2$ \\
$bc$ binding & ${-}170.8$ & $bc$ binding & {-}159.4 \\
Total & $7133.7\pm 13$ & Total & $7255.3 \pm 13$ \\ \hline \hline
\end{tabular}
\strut\vskip-0.7cm\strut
\end{center}
\end{table}

\begin{table}
\caption{\small%
Contributions to the mass of $\Xi_{cc}$, the spin-1/2 ground state
of $ccq$ ($q= u,d$); isospin breaking ignored.
\label{tab:compxi}}
\begin{center}
\begin{tabular}{c r c r} \hline \hline
\multicolumn{2}{c}{Baryonic quarks} & \multicolumn{2}{c}{Universal quarks} \\
Contribution & Value (MeV)  & Contribution & Value (MeV) \\ \hline
    --   &   --    &  $S$  &  165.1 \\
$2m^b_c+m^b_q$ & 3783.9 & $2m_c+m_q$ & 3619.7 \\
$a_{cc}/(m^b_c)^2$ & 14.2 & $a_{cc}/(m_c)^2$ & 14.2 \\
${-}3a/(m^b_q m^b_c))$ & ${-}42.4$ & ${-}3a/(m_qm_c)$ & ${-}42.4$ \\
$cc$ binding & ${-}129.0$ & $cc$ binding & ${-}121.3$ \\
Total &$3627\pm12$&Total &$3635\pm12$  \\ \hline \hline
\end{tabular}
\end{center}
\end{table}

To summarize:  The exotic tetraquark reported
by LHCb \cite{LHCbTalk:2020} weighs in favor of the string-junction picture
with universal quark masses for mesons and baryons.  The observed $0^+$ mass
is $2866 \pm 5$ MeV, while we predict $2754 \pm 12$ MeV in the baryonic-quark
scheme and $2863 \pm 12$ MeV in the string-junction (universal quark mass)
picture.

The compact tetraquark picture makes a clear prediction that $X_0(2900)$ is
an isoscalar. This prediction can be tested, e.g., by looking for it
charged partner in the $D^0 K^-$ invariant mass in $B^- \to \bar D^0 D^0 K^-$.

The bottom analogue of $X_0(2900)$, with quark content 
$(bs\bar u\bar d)$, is predicted at $6213\pm 12$ MeV.

There remains the question of how to interpret the broader $1^-$ peak
(\ref{eqn1}).  One possibility, as mentioned, is an artifact due to
rescattering at and above the $D^*K^*$ threshold.  While the current LHCb
analysis \cite{LHCbTalk:2020} prefers a $1^-$ assignment, it is worth
mentioning that if this peak is really due to a $J^P = 2^+$ resonance, it would
also populate the Dalitz plot band at 2.9 GeV non-uniformly along its length.
Such a state could be a $D^* K^*$ molecule, very close to the threshold at
2902 MeV, with pion exchange providing a major part of the attraction.
On the other hand, the lighter and narrower $0^+$ state 
is 36 MeV below the $D^*K^*$ threshold, which is 
much too large for binding energy in a hadronic molecule.

The predictions for masses of the $bb \bar u \bar d,~cc \bar u \bar d$, and $bc
\bar u \bar d$ masses are shifted upward in the string-junction picture by
126, 118, and 122 MeV, respectively.  The $bb \bar u \bar d$ state is still
stable with respect to strong and EM interactions, as its mass is
predicted to lie 89 MeV below threshold for strong decay and 44 MeV below that
for radiative decay, while the $cc \bar u \bar d$ and $bc \bar u \bar d$ masses
lie well above strong decay thresholds.

The prediction of the doubly charmed baryon mass \cite{Karliner:2014gca} based
on baryonic quarks, $M(\Xi_{cc}) = 3627\pm12$ MeV, is only raised by 8 MeV
in the string-junction picture with universal quark masses, still compatible
with the experimental value of $(3621.55 \pm 0.23 \pm 0.30)$ MeV.

Further tests of the physical picture discussed here will become possible
when additional tetraquark states are observed experimentally. We hope that
this will be possible in the foreseeable future.

In particular, since the doubly charmed baryon $\Xi_{cc}^{++}~(ccu)$
has been observed by LHCb \cite{Aaij:2017ueg,Aaij:2019uaz}, the 
data accumulated so far might make it possible to observe the 
$T(cc\bar u\bar d)$ tetraquark decaying into $D^0 D^{*+}$ and $D^+ D^{*0}$.
The measured mass will then tell us whether the string-junction
description applies to this state as well, or is the physical picture of
Ref.~\cite{Karliner:2017qjm} more appropriate for a tetraquark with two
truly heavy quarks.

We thank Alex Bondar, Tim Gershon, and Tomasz Skwarnicki for useful discussions
about the data presented by LHCb in the seminar~\cite{LHCbTalk:2020}; Lance
Dixon, Gabriele Veneziano and Jian-Rong Zhang for comments on the manuscript;
and Eulogio Oset for pointing out Ref.\ \cite{Molina:2010tx}.
The research of M.K. was supported in part by NSFC-ISF grant No.\ 0603219411.

\vfill\eject

Note added: After this work was posted on the arXiv, it was pointed out to us
that a prediction was made \cite{Molina:2010tx} using coupled-channel unitarity
of a $D^{*+}K^{*-}$ bound state with $C=1$, $S=-1$, $J^P=0^+$, 
with pole mass of 2848 MeV (i.e., binding energy of 54 MeV) and width between 23
MeV and 59 MeV, depending on the value of a cutoff parameter.

\end{document}